\documentstyle[psfig,12pt]{article} 
\def\lsim{\:\raisebox{-0.5ex}{$\stackrel{\textstyle<}{\sim}$}\:}

\def\be{\begin{equation}}      
\def\ee{\end{equation}}
\def\bear{\be\begin{array}}
\def\eear{\end{array}\ee}
\def\bea{\begin{eqnarray}}
\def\eea{\end{eqnarray}}

\def\21{$SU(2) \ot U(1)$}
\def\ot{\otimes}
\def\ie{{\it i.e.}}

\def\half{{\textstyle{1 \over 2}}}

\def\quarter{{\textstyle{1 \over 4}}}
\def\fourth{{\textstyle{1 \over 4}}}

\def\bold#1{\setbox0=\hbox{$#1$}
     \kern-.025em\copy0\kern-\wd0
     \kern.05em\copy0\kern-\wd0
     \kern-.025em\raise.0433em\box0 }
\begin{document}
\begin{titlepage}
\begin{flushright}
hep-ph/9909212\\
FTUV/99-58\\
IFIC/99-61\\
FSU-HEP-990710\\
UCCHEP/2-99\\
%%FSU--HEP--990710\\
September 1999
\end{flushright}
\vspace*{5mm}
\begin{center}
{\Large \bf Supersymmetry Unification Predictions for
$\bold{m_{top}}$, $\bold{V_{cb}}$ and $\bold{\tan\beta}$}\\[15mm]
{\large
M. A. D\'\i az${}^{2,3}$, J. Ferrandis${}^1$, and J. W. F.
Valle${}^1$}\\
\hspace{3cm}\\
{\small ${}^1$Departamento de F\'\i sica Te\'orica, IFIC-CSIC, 
Universidad de Valencia}\\ 
{\small Burjassot, Valencia 46100, Spain}
\vskip 0.1cm
{\small ${}^2$High Energy Group, Physics Department 
Florida State University}\\ 
{\small Tallahassee, Florida 32306, USA}
\vskip 0.1cm
{\small ${}^3$Facultad de F\'\i sica, 
Universidad Cat\'olica de Chile}\\
{\small Av. Vicu\~na Mackenna 4860, Santiago, Chile}
\hspace{3cm}\\
\end{center}
\vspace{5mm}
\begin{abstract}

We study the predictions for $m_{top}$, $\tan\beta$ and $V_{cb}$ in a
popular texture ans\"atze for the fermion mass matrices. We do this
both for the Minimal Supersymmetric Standard Model (MSSM) and for the
simplest model (MSSM--BRpV) where a bilinear R--Parity violating term
is added to the superpotential.  We find that taking the experimental
values for $m_{top}$ and $V_{cb}$ at $99\%$ c.l.~and the GUT relations
$h_b=h_{\tau}$ and $V_{cb}^2=h_c/h_t$ within $5\%$, the large
$\tan\beta$ solution, characteristic in the MSSM with bottom--tau
unification, becomes disallowed. In contrast the corresponding allowed
region for the MSSM--BRpV is slightly larger. We also find that
important modifications occur if we relax the texture conditions at
the GUT scale. For example, if the GUT relations are imposed at
$40\%$, the large $\tan\beta$ branch in the MSSM becomes fully
allowed.  In addition, in MSSM--BRpV the whole $\tan\beta-m_{top}$
plane become allowed, finding unification at any value of $\tan\beta$.

\end{abstract}

\end{titlepage}

\setcounter{page}{1}

\section{Introduction}

Grand unified (GUT) symmetries \cite{GG} combined with flavour
symmetries 
\cite{texrev} constitute the most promising way of understanding the
structure of flavour masses and mixings. These masses and mixings
constitute the majority of the unknown parameters of the Standard
Model (SM). On the other hand, supersymmetry allows the unification of
gauge couplings to succeed where the SM fails \cite{GUT,gaugeUnif},
implying the prediction of one of the three gauge coupling constants.

In some GUT models [for example SU(5)], the bottom quark and the tau 
lepton Yukawa couplings are equal at the unification scale, and the 
predicted ratio $m_b/m_{\tau}$ at the weak scale agrees with
experiments.
Several studies have been made about the effect of supersymmetry on
gauge and Yukawa unification. In the Minimal Supersymmetric Standard
Model
(MSSM) bottom--tau unification is achieved at two disconnected and small 
regions of $\tan\beta$ (the ratio of the two vacuum expectation values), 
one at small and the other at large $\tan\beta$ \cite{BBO,CPW,MSSMYuk}.
 
Recently it was shown that if to the MSSM we add Bilinear R--Parity
Violation (BRpV) \cite{e3othersa,e3othersb,epsrad,v3cha}, the
unification of the bottom and tau Yukawa couplings at the scale
$M_{GUT}\approx 10^{16}$ GeV (where the gauge couplings unify) is
dramatically different from the MSSM \cite{YukUnif}. In the BRpV case,
bottom--tau unification is achieved at any value of $\tan\beta$
provided the vacuum expectation value $v_3$ of the tau sneutrino is
chosen appropriately. In addition, it was shown that the prediction of
$\alpha_s$, which in the MSSM is $2\sigma$ too high, in BRpV can be
lowered by more than $1\sigma$ with respect to the MSSM prediction and
therefore can lie closer to the experimental measurement \cite{alfas}.

The study of BRpV is motivated by the fact that it provides a simple
and useful parametrization of many of the features of a class of
models in which R-Parity is spontaneously broken \cite{SponRpB}. One
of the main features of R--Parity violating models is the appearance
of masses for the neutrinos \cite{SponRpB,rpre}, attracting a lot of
attention~\cite{rprecent} since the latest results from
Super--Kamiokande~\cite{Fukuda}. It has in fact been demostrated that
this model offers an attractive and predictive scheme for neutrino
masses and mixing parameters which accounts for the observed data from
atmospheric and solar neutrino observations~\cite{Romao:1999up}.

In this paper we update the analysis of the relations between
$m_{top}$ and $\tan\beta$ within the MSSM for the case in which the
bottom and tau Yukawa couplings unify and using the CKM matrix element
$V_{cb}$ that follows from the simplest Yukawa texture, adopting the
most recent experimental values $0.036<|V_{cb}|<0.042$ at $90\%$
c.l. prescribed by the Particle Data Group \cite{PDB}. In addition,
following closely the method presented in ref.~\cite{YukUnif} we
repeat the analysis for the MSSM--BRpV model~\cite{epsrad,v3cha} and
compare the results obtained with those found in the MSSM.

\section{Zero Texture Ans\"atze}

Flavour symmetries in two and three generations were first proposed 
in \cite{WZ_F}. The validity of such mass matrix ans\"atze at the GUT 
scale was postulated by \cite{GJ} and later the ans\"atze was modified 
in \cite{GN}. The final version of the mass matrix we are considering 
here is given in \cite{DHR} and corresponds to
\begin{equation}
{\bf{h_U}}=\left[\matrix{0 & C & 0 \cr C & 0 & B \cr 0 & B & A}\right]
\,,\qquad
{\bf{h_D}}=\left[\matrix{0 & Fe^{i\phi} & 0 \cr Fe^{-i\phi} & E & 0 
\cr 0 & 0 & D}\right]
\,,\qquad
{\bf{h_E}}=\left[\matrix{0 & F & 0 \cr F & -3E & 0 \cr 0 & 0 & D}\right]
\label{ansaetze}
\end{equation}
where ${\bf{h_U}}$, ${\bf{h_D}}$, and ${\bf{h_E}}$ are the up--type
quark, down--type quark, and charged lepton Yukawa matrices
respectively. The dimention-less parameters $A$, $B$, $C$, $D$, $E$,
and $F$ are real and $\phi$ is the only phase.

The fact that the third diagonal matrix element in the down--type quark 
and the charged lepton Yukawa matrices are the same indicates
bottom--tau 
unification at the GUT scale. Another interesting prediction refers to 
the CKM matrix element $V_{cb}$. After defining running CKM matrix 
elements \cite{BBO}, the following relation holds at the GUT scale
\begin{equation}
\Big|V_{cb}(M_{GUT})\Big|=\sqrt{{h_c(M_{GUT})}\over{h_t(M_{GUT})}}
\label{VcbGUT}
\end{equation}
In this way, together with the bottom--tau unification condition
\begin{equation}
h_b(M_{GUT})=h_{\tau}(M_{GUT})\,,
\label{hbhtauGUT}
\end{equation}
The corresponding relations between $m_{top}$, $V_{cb}$ and
$\tan\beta$ have been derived in the literature~\cite{BBO,ARDH}.  Here
we closely followed the method developed in \cite{BBO}, updating the
analysis of these relations for the case in which bottom--tau Yukawa
couplings unify, as indicated by eq.~(\ref{hbhtauGUT}), and with the
CKM matrix element $V_{cb}$ given by eq.~(\ref{VcbGUT}) and satisfying
the experimental constraint at the weak scale $0.036<|V_{cb}|<0.042$
at $90\%$ c.l. \cite{PDB}. This is done first for the MSSM case. In
addition, following closely ref.~\cite{YukUnif} we do the same
analysis for the MSSM--BRpV model~\cite{epsrad,v3cha}.

\section{Bilinear R--Parity Violation}

The MSSM--BRpV has one bilinear term in the superpotential for each
generation. In this way, after including one-loop radiative
corrections, neutrino masses and mixings can be predicted
\cite{Romao:1999up}.  For our present purposes in this paper
it will sufficient to consider lepton and Rp violation only in the tau
sector. In this case, the superpotential has the following bilinear
terms
\begin{equation} 
W_{Bi}=\varepsilon_{ab}\left[
-\mu\widehat H_d^a\widehat H_u^b
+\epsilon_3\widehat L_3^a\widehat H_u^b\right]\,,
\label{eq:WBi}
\end{equation}
with $\mu$ and $\epsilon_3$ having units of mass. The MSSM
superpotential is recovered if we take $\epsilon_3=0$. The BRpV term
can disappear from the superpotential if we make the rotation defined
by $\mu'\widehat H_d'=\mu\widehat H_d-\epsilon_3\widehat L_3$ and
$\mu'\widehat L_3'=\epsilon_3\widehat H_d+\mu\widehat L_3$, with
$\mu'^2=\mu^2+\epsilon_3^2$. Nevertheless, BRpV effects are
reintroduced through the soft terms in such a way that sneutrino
vacuum expectation values are present in both basis:
$\langle\widetilde L_3\rangle=v_3/\sqrt{2}$ and $\langle\widetilde
L'_3\rangle=v'_3/\sqrt{2}$. The VEV $v_3$ contributes to the $W$ boson
mass according to $m_W^2=\quarter g^2(v_d^2+v_u^2+v_3^2)$.  On the
other hand, the relations of quark masses with Yukawa couplings are
the same in BRpV--MSSM as in the MSSM, namely
\begin{equation}
h_{t,c}^2={{2m_{t,c}^2}\over{v_u^2}}\,,\qquad
h_b^2={{2m_b^2}\over{v_d^2}}\,.
\label{QuarkYuk}
\end{equation}
except for the numerical value of $v_d$.

However, in the BRpV model the tau lepton mixes with the charginos,
and in the {\sl{original}} basis where
$\psi^{+T}=(-i\lambda^+,\widetilde H_u^1,\tau_R^+)$ and
$\psi^{-T}=(-i\lambda^-,\widetilde H_d^2,\tau_L^-)$, the charged
fermion mass terms in the Lagrangian are ${\cal L}_m=-\psi^{-T}{\bold
M_C}\psi^+$, with the mass matrix given by
\begin{equation} 
{\bold M_C}=\left[\matrix{ 
M & {\textstyle{1\over{\sqrt{2}}}}gv_u & 0 \cr 
{\textstyle{1\over{\sqrt{2}}}}gv_d & \mu &  
-{\textstyle{1\over{\sqrt{2}}}}h_{\tau}v_3 \cr 
{\textstyle{1\over{\sqrt{2}}}}gv_3 & -\epsilon_3 & 
{\textstyle{1\over{\sqrt{2}}}}h_{\tau}v_d}\right] 
\label{eq:ChaM3x3} 
\end{equation} 
where $M$ is the $SU(2)$ gaugino mass. In the limit $\epsilon_3=v_3=0$
the
MSSM chargino mass matrix is recovered in the upper--left $2\times 2$
sub-matrix and at the same time the tau mass relation in the third
diagonal
element [analogous to the bottom mass relation in eq.~(\ref{QuarkYuk})].
This tau mass relation is no longer valid in BRpV--MSSM and it is 
modified to
\begin{equation}
h^2_{\tau}={{2m_{\tau}^2}\over{v_d^2}}{1\over{1+\delta}}
\,,\qquad
\delta={{v_3^2}\over{v_d^2}}+\left[
{{(A-m_{\tau}^2)\mu'^2}\over{Tm_{\tau}^2-m_{\tau}^4-\Delta}}\right]
{{v'^2_3}\over{v_d^2}}
\label{htaudelta}
\end{equation}
where $A$, $T$, and $\Delta$ refer to the upper left $2\times 2$
sub-matrix 
of the $3\times 3$ matrix $\bold{M'^T_C M'_C}$: $A$ is its first
diagonal
element, $T$ is its trace, and $\Delta$ is its determinant. The matrix
$\bold{M'_C}$ is the chargino mass matrix analogous to 
eq.~(\ref{eq:ChaM3x3}) but in the {\sl{rotated}} basis. It is easy to
see 
that $\bold{M_C}\rightarrow\bold{M'_C}$ when 
$(\mu,\epsilon_3,v_d,v_3)\rightarrow(\mu',0,v'_d,v'_3)$.

\section{RGE's and Matching Conditions}

We use two-loop MSSM RGE's at scales $Q>M_{SUSY}$ and two loop SM RGE's
at scales $Q<M_{SUSY}$. Therefore, we include leading and 
next--to--leading logarithmic supersymmetric threshold corrections in 
the approximation where all the SUSY particles decouple at the same 
scale $Q=M_{SUSY}$. In this way, the matching conditions at $Q=M_{SUSY}$ 
are defined by the continuity of the quark and lepton running masses at 
that scale, which translates into matching conditions on Yukawa 
couplings given in MSSM--BRpV as
\begin{eqnarray}
&&\lambda_{t,c}(M_{SUSY}^-)=h_{t,c} (M_{SUSY}^+) \sin\beta\sin\theta \,,
\nonumber\\
&&\lambda_b(M_{SUSY}^-)=h_b (M_{SUSY}^+) \cos\beta\sin\theta \,,
\label{BounYuk}\\
&&\lambda_{\tau}(M_{SUSY}^-)=h_{\tau} (M_{SUSY}^+) \cos\beta\sin\theta
\sqrt{1+\delta}\,,
\nonumber
\end{eqnarray}
where we have defined the angles $\beta$ and $\theta$ according to 
spherical coordinates
\begin{equation}
v_d=v\cos\beta\sin\theta\,,\qquad v_u=v\sin\beta\sin\theta\,,\qquad
v_3=v\cos\theta\,,
\label{vevdef}
\end{equation}
with $v=246$ GeV. Note that the MSSM relation $\tan\beta=v_u/v_d$ is 
preserved.
In addition, the boundary condition for the quartic Higgs coupling is
given by 
\begin{equation}
\qquad\lambda(M_{SUSY}^-) = \quarter
\Big[(g^2(M_{SUSY}^+)+g'^2(M_{SUSY}^+) \Big] (\cos2\beta\sin^2\theta+
\cos^2\theta)^2\,.
\label{BounHiggsCoup}
\end{equation}
The corresponding MSSM boundary conditions are obtained by setting 
$\theta=\pi/2$.

Starting at the scale $Q=m_Z$ we randomly vary the parameters 
$\alpha^{-1}_{em}(m_Z) = 128.896 \pm0.090$, $\sin^2\theta_w(m_Z) =
0.2322 \pm 0.0010$, and $\alpha_s(m_Z)=0.118 \pm 0.003$
\cite{LEPinternal},
looking for solutions with gauge unification at a scale $M_{GUT}$ with
a common gauge coupling $\alpha_{GUT}$. These solutions are concentrated 
in a region of the plane $M_{GUT}-\alpha_{GUT}$ centered around 
$M_{GUT}\approx 2.3 \times10^{16}$ GeV and 
${\alpha_{GUT}}^{-1} \approx 24.5$. For simplicity, from now on, we fix 
the unification scale to that value. Since $M_{GUT}$ depends 
on other input parameters, this simplification implies that we don't
have
``perfect'' unification throughout our sampling. Nevertheless, we have 
checked that unification is good up to $0.4\%$.

Next, we evolve the Yukawa couplings using two-loop RGEs, starting
from the experimental values of the quark and lepton masses at the
weak scale and imposing unification of bottom--tau Yukawa couplings at
$M_{GUT}$ within $5\%$. Matching conditions at $M_{SUSY}$ are well
known in the MSSM. The main difference in our BRpV model lies in the
fact that since the sneutrino vacuum expectation value $v_3$
contributes also to the $W$--boson mass, the Higgs VEVs will be in
general smaller. This in turn makes the down-type Yukawa couplings
larger than in the MSSM. In addition, tau mixing with charginos makes
the tau Yukawa coupling $h_{\tau}$ a quantity which not only depends
on $\tan\beta$ but also on the other chargino ($M$, $\mu$) and BRpV
($\epsilon_3$, $v_3$) parameters~\cite{v3cha}.

Apart from imposing unification of bottom--tau Yukawa couplings we
calculate the texture prediction for $V_{cb}$ at the weak scale with
the boundary condition given in eq.~(\ref{VcbGUT}) within
$5\%$. Regarding the MSSM part of the analysis, we have updated the
analysis in refs.~\cite{BBO} and \cite{ARDH} by incorporating the most
recent experimental values of the top quark mass and of $V_{cb}$. In
contarst, in the case of the the BRpV model, the analysis is done for
the first time.

\section{Numerical Results}

\begin{figure}
\centerline{\psfig{file=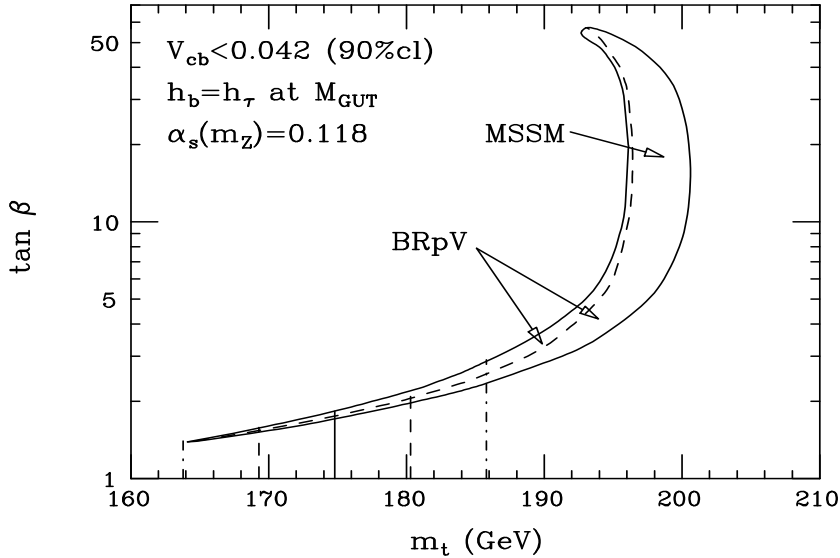,height=9cm,angle=90,width=0.9\textwidth}}
\caption{Allowed regions in the $\tan\beta-m_{top}$ plane where 
bottom--tau Yukawa unification is possible together with the texture
prediction for $V_{cb}$. Accepted values of $V_{cb}$ lie in the $90\%$
c.l.
The vertical lines correspond to the experimental measurement of the top
quark mass, with its central value (solid), $1\sigma$ (dashes), and
$2\sigma$ (dot--dash) regions.}
\label{tl90}
\end{figure}
\begin{figure}
\centerline{\psfig{file=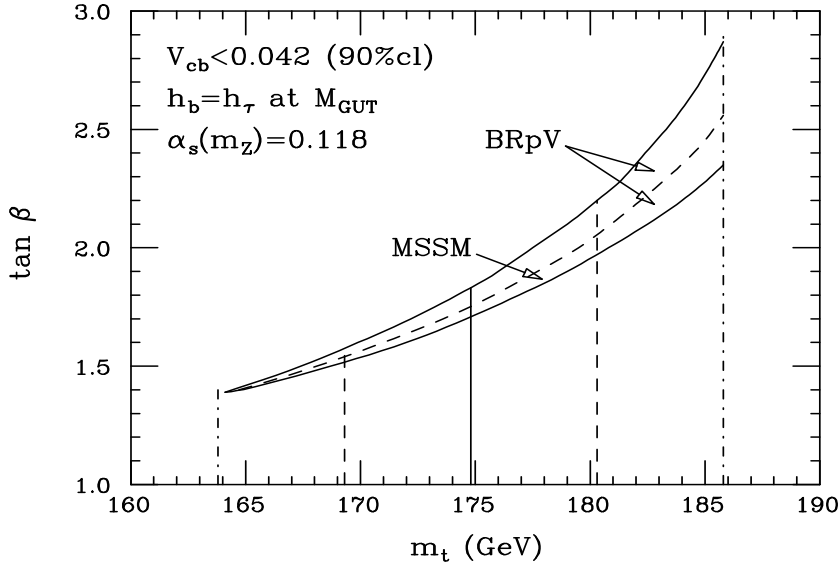,height=9cm,angle=90,width=0.9\textwidth}}
\caption{Allowed regions in the $\tan\beta-m_{top}$ plane where 
bottom--tau Yukawa unification is possible together with the texture
prediction for $V_{cb}$. This is a magnification of the low $\tan\beta$
region displayed in the previous figure. Accepted values of $V_{cb}$ 
lie in the $90\%$ c.l.}
\label{ts90}
\end{figure}
In Fig.~\ref{tl90} we display the regions in the $\tan\beta-m_{top}$
plane where bottom--tau Yukawa unification occurs together with the
prediction for the Cabibbo--Kobayashi--Maskawa matrix element
$V_{cb}$. This prediction lies in the region indicated by experiment,
\ie, $0.036<V_{cb}<0.042$, at $90\%$ c.l. Nevertheless, it is worth 
mentioning that we do not find any point with $V_{cb}<0.039$. The
space between the solid curves is the allowed region in
MSSM--BRpV. Similarly, the space between the right solid curve and the
dashed curve is the allowed region in the MSSM.  The solid curve at
the right is common to both models and corresponds to the Landau pole
of the quark Yukawa couplings (quasi--fixed point). The solid vertical
line corresponds to the central value of the experimental measurement
for $m_{top}$, and the dashed (dot--dashed) lines are the $1\sigma$
($2\sigma$) limits.

It is known that bottom--tau unification in the MSSM is obtained in a
region similar to the one in Fig.~\ref{tl90} but including an extra
branch at high $\tan\beta$. By imposing the texture prediction for
$V_{cb}$ this branch disappears.  It can be observed from the figure
that the MSSM--BRpV region is only slightly larger than the MSSM
region. Nevertheless, in the $2\sigma$ region for the top quark mass
the MSSM--BRpV allowed region is about twice as large as the MSSM
one. This can be seen in Fig.~\ref{ts90} which is a blow up of the
previous figure. However in our scan we did not find any solution in
the large $\tan\beta$ branch within the MSSM nor the MSSM--BRpV.

\begin{figure}
\centerline{\psfig{file=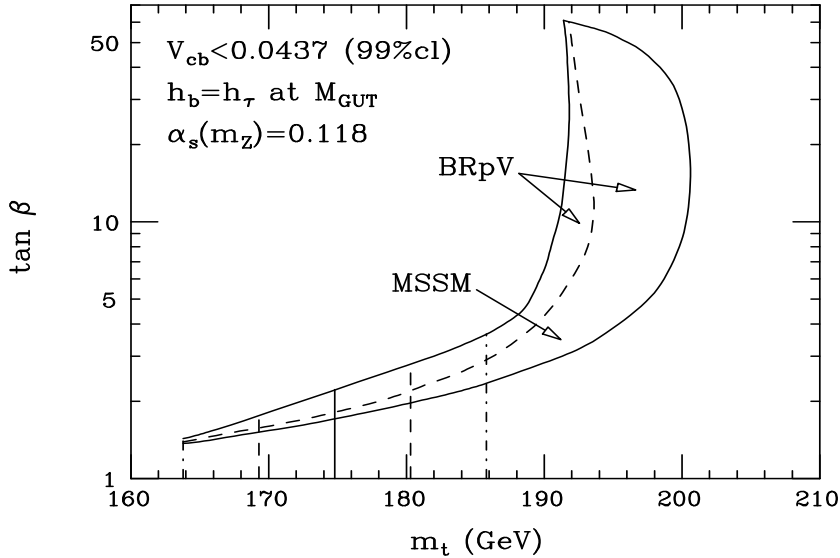,height=9cm,angle=90,width=0.9\textwidth}}
\caption{Allowed regions in the $\tan\beta-m_{top}$ plane where 
bottom--tau Yukawa unification is possible together with the texture
prediction for $V_{cb}$. Accepted values of $V_{cb}$ lie in the $99\%$
c.l.}
\label{tl99}
\end{figure}
\begin{figure}
\centerline{\psfig{file=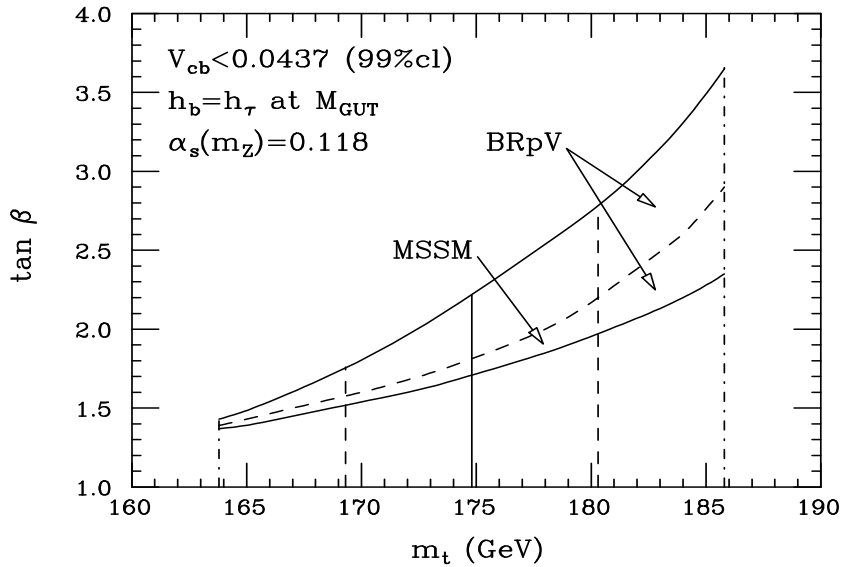,height=9cm,angle=90,width=0.9\textwidth}}
\caption{Allowed regions in the $\tan\beta-m_{top}$ plane where 
bottom--tau Yukawa unification is possible together with the texture
prediction for $V_{cb}$. This is a magnification of the low $\tan\beta$
region displayed in the previous figure. Accepted values of $V_{cb}$ 
lie in the $99\%$ c.l.}
\label{ts99}
\end{figure}

In Fig.~\ref{tl99} we have relaxed the allowed values of $V_{cb}$ at
the weak scale. In this figure we consider $V_{cb}<0.0437$ which
naively corresponds to the $99\%$ c.l.~region (here we don't find
solutions with $V_{cb}<0.039$ neither). Although the allowed regions
are bigger, the large $\tan\beta$ branch is still not present.
Nevertheless, the difference between the MSSM--BRpV and the MSSM is
more pronounced in this case, as it can be seen from Fig.~\ref{ts99}
where we blow up the region compatible with the top quark mass
measurement.  Note that preliminary results of Higgs searches by the
ALEPH collaboration \cite{aleph:1999} which rule out low values of
$\tan\beta$, pushing $m_t$ to high values in the MSSM would not
necessarily hold in our BRpV case, due to the importance of novel
Higgs boson decay channels \cite{e3othersa}.  

The previous four figures have been obtained imposing the validity of
the bottom--tau Yukawa unification condition in eq.~(\ref{hbhtauGUT})
and the $V_{cb}$ texture condition in eq.~(\ref{VcbGUT}) at the $5\%$
level. In the next two figures we explore the effect of relaxing the
$5\%$. As we can see, the effect is very interesting.

\begin{figure}
\centerline{\psfig{file=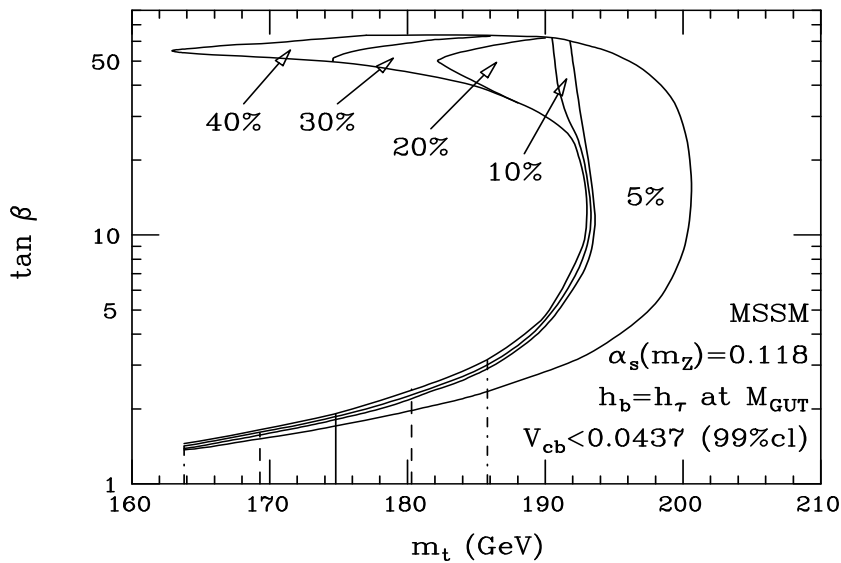,height=9cm,angle=90,width=0.9\textwidth}}
\caption{Allowed regions in the $\tan\beta-m_{top}$ plane for the MSSM.
The 
texture conditions at the GUT scale are relaxed to lie within the
indicated percent level.  Accepted values of $V_{cb}$ at the weak
scale lie in the $99\%$ c.l.}
\label{tl99Mr}
\end{figure}
\begin{figure}
\centerline{\psfig{file=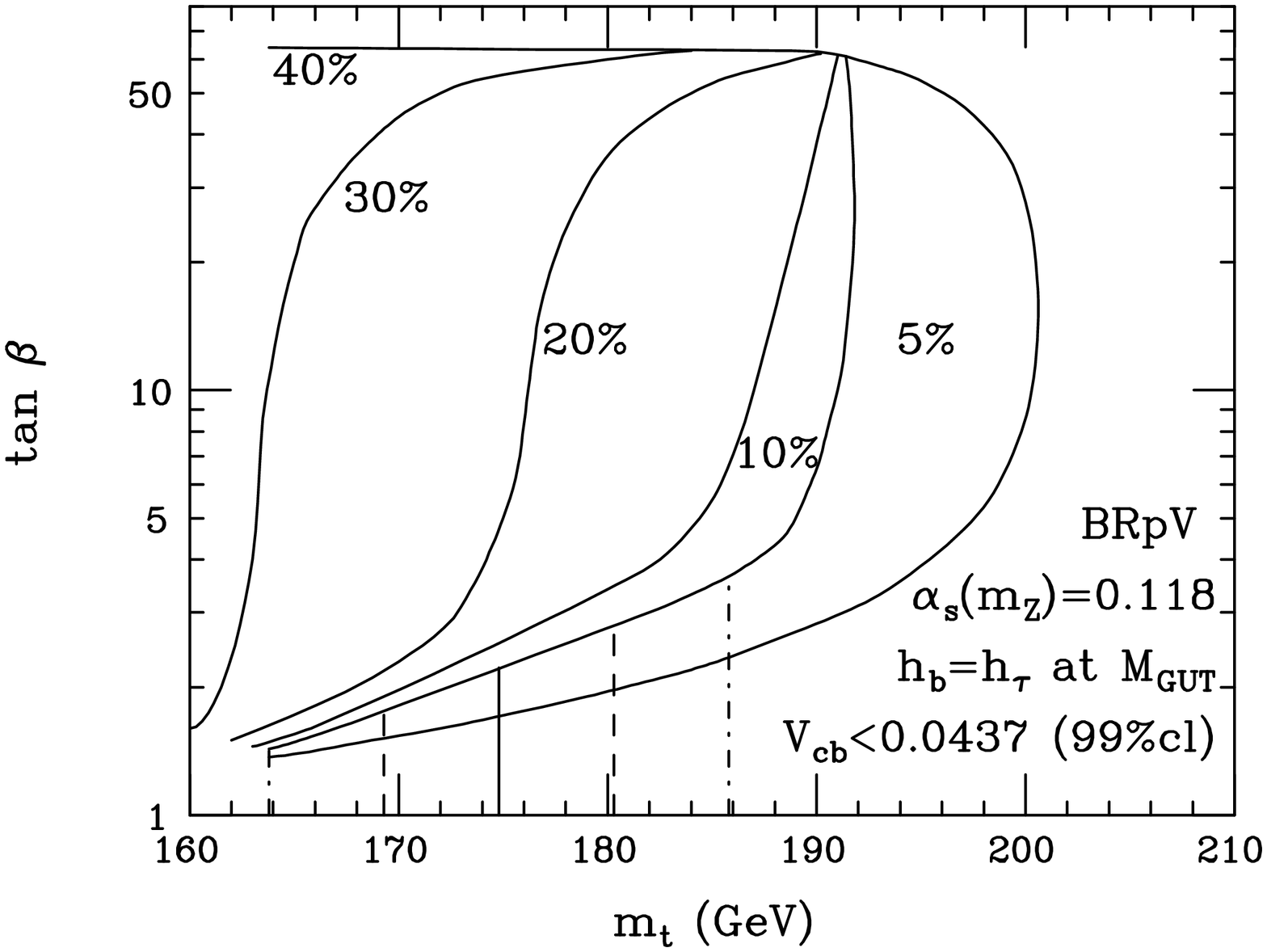,height=9cm,angle=90,width=0.9\textwidth}}
\caption{Allowed regions in the $\tan\beta-m_{top}$ plane for the
MSSM--BRpV. 
The texture conditions at the GUT scale are relaxed to lie within the
indicated percent level.  Accepted values of $V_{cb}$ at the weak
scale lie in the $99\%$ c.l.}
\label{tl99Br}
\end{figure}
In Fig.~\ref{tl99Mr} we have plotted the allowed regions in the
$\tan\beta-m_{top}$ plane within the MSSM. There are five regions each
one labelled by the maximum deviation in percent accepted for the
conditions in eqs.~(\ref{VcbGUT}) and (\ref{hbhtauGUT}). Clearly, the
large $\tan\beta$ branch of the MSSM slowly reappears as we relax the
GUT conditions and it is fully present in the $40\%$ case. Therefore
in this case two solutions are possible, one at large and one at small
values of $\tan\beta$, in order to account for the measurement of the
top quark mass. The situation is different in MSSM--BRpV. In this case
the whole interval for $\tan\beta$ compatible with perturbativity of
Yukawa couplings slowly reappears as we relax the GUT conditions. If
we accept the GUT conditions within $40\%$, then the allowed region is
all the space at the left of the quasi--infrared fixed curve. In this
case, the prediction for $V_{cb}$, $m_{top}$ and $\tan\beta$ in BRpV
is dramatically different from that in the MSSM. This was already
pointed out for bottom--tau Yukawa unification in ref.~\cite{YukUnif}.

\section{Discussion}
%: $\bold{b-\tau}$ Unification, $\bold{V_{cb}}$, and 
%$\bold{\nu_{\tau}}$ Mass}

In this section we provide a way to understand of the results
presented above in the figures 1 to 6. In to do this we make some
approximations so that the relevant RGE's have simple analytical
solutions. First of all, let us consider the question of why in BRpV
bottom--tau Yukawa unification is achieved at any value of
$\tan\beta$, as opposed to the MSSM, where only two disconnected
regions of $\tan\beta$ are allowed
\cite{YukUnif}. We notice first that the quark and lepton masses are
related to the different VEVs and Yukawa couplings in the following
way
\begin{equation}
m_{top}^2=\half h_t^2v_u^2\,,\qquad m_b^2=\half h_b^2v_d^2\,,\qquad
m_{\tau}^2=\half h_{\tau}^2v_d^2(1+\delta)\,,
\label{fermmass}
\end{equation}
where $\delta$ depends on the parameters of the chargino/tau mass matrix
and is positive \cite{v3cha,YukUnif}. This implies that the ratio of 
the bottom and tau Yukawa couplings at the weak scale is given by
\begin{equation}
{{h_b}\over{h_{\tau}}}(m_{weak})={{m_b}\over{m_{\tau}}}\sqrt{1+\delta}
\label{ratio1}
\end{equation}
and grows as $|v_3|$ is increased.

On the other hand, if $h_b$ and $h_{\tau}$ unify at the GUT scale, then 
at the weak scale its ratio can be approximated by
\begin{equation}
{{h_b}\over{h_{\tau}}}(m_{weak})\approx exp\left[
{1\over{16\pi^2}}\left({16\over3}g_s^2-3h_b^2-h_t^2\right)
\ln{{M_{GUT}}\over{m_{weak}}}\right]
\label{hb_htau}
\end{equation}
implying that the combination $3h_b^2+h_t^2$ should decrease when
$|v_3|$
increases.

In the MSSM region of high $\tan\beta$ the bottom quark Yukawa coupling
dominates over the top one, and the opposite happens in the region of
low $\tan\beta$. Therefore, at high (low) values of $\tan\beta$, the
Yukawa coupling $h_b$ ($h_t$) will decrease if $|v_3|$ increases, which
implies an increase of $v_d$ ($v_u$) in order to keep constant the quark
masses. Similarly, in order to keep constant the $W$ mass,
$m_W^2=\fourth g^2(v_d^2+v_u^2+v_3^2)$, the VEV $v_u$ ($v_d$) decreases
at the same time. This implies that unification occurs at lower (higher)
values of $\tan\beta$ as $|v_3|$ increases. This explains why in BRpV
intermediate values of $\tan\beta$ are compatible with bottom--tau
unification.

Let us now understand why the high $\tan\beta$ branch is not allowed
when we impose the $|V_{cb}|$ constraint at the unification scale. The
RGE for the CKM angle $|V_{cb}|$ is \cite{BBO}
\begin{equation}
{{d|V_{cb}|}\over{dt}}=-{{|V_{cb}|}\over{16\pi^2}}\Big(h_t^2+h_b^2\Big)
\label{VcbRGE}
\end{equation}
where $t=\ln(Q)$. In addition, the RGE for the ratio between the charm 
and top quark Yukawa couplings $R_{c/t}\equiv h_c/h_t$ is
\begin{equation}
{{dR_{c/t}}\over{dt}}=-{{R_{c/t}}\over{16\pi^2}}\Big(3h_t^2+h_b^2\Big)\,.
\label{RctRGE}
\end{equation}
Imposing now the relation in eq.~(\ref{VcbGUT}) at the GUT scale, we
obtain at the weak scale
\begin{equation}
{{R_{c/t}}\over{|V_{cb}|^2}}(m_{weak})\approx exp\left[{1\over{16\pi^2}}
\Big(h_t^2-h_b^2\Big)\ln{{M_{GUT}}\over{m_{weak}}}\right]
\label{RctVcbWeak}
\end{equation}
where we have approximated the RGE's to first order in perturbation
series.  Since the left hand side of eq.~(\ref{RctVcbWeak}) is greater
than one (approximately equal to 1.5), it is clear that the GUT
condition $R_{c/t}=|V_{cb}|^2$ prefers the region of parameter space
where the top Yukawa coupling is large while the bottom Yukawa
coupling is small. This is obtained at small values of $\tan\beta$,
since our definition of $\tan\beta=v_u/v_d$ retains the MSSM relation
$h_b/h_t=m_bt_{\beta}/m_{top}$ \footnote{ In
refs.~\cite{alfas,ferrandis} it was defined as
$\tan\beta'=v_u/\sqrt{v_d^2+v_3^2}$ which has the advantage of being
invariant under rotations defined at the beginning of section 3, but
spoils the relation between $h_t$ and $h_b$ described in the text.  }.
This explains what it is seen in Figs.~\ref{tl99Mr} and \ref{tl99Br}.

If the GUT conditions are relaxed to more than 5\%,
eq.~(\ref{RctVcbWeak})
should be modified by adding a numerical factor different from one in 
front of the exponential. The effect is to allow larger values of $h_b$
that can only be achieved in BRpV by increasing $v_3$ without having to
go 
to very large values of $\tan\beta$ as in the MSSM. Consequently, the
plane 
$m_{top}-\tan\beta$ is filled up in BRpV and not in the MSSM.

Now we would like to understand why in BRpV larger values of $\tan\beta$
are acceptable compared with the MSSM when imposing the GUT conditions
at 
5\%. This effect is observed in Figs.~\ref{tl90} to \ref{ts99}. We
notice 
first that our numerical results with 5\% of unification indicate that 
BRpV accepts values of $h_t$ slightly smaller than the MSSM (the upper 
bound on $h_t$ is the same in both models). Considering the base 
independent parameter $\cos\chi$ defined for example in 
refs.~\cite{alfas,ferrandis} and whose expression in our basis is 
$\cos\chi=v_d/\sqrt{v_d^2+v_3^2}$, we have for the top quark Yukawa 
coupling
\begin{equation}
h_t^2={{g^2m_{top}^2}\over{2m_W^2}}\Bigg(1+{1\over{t_{\beta}^2c_{\chi}^2}}
\Bigg)\,.
\label{htopchi}
\end{equation}
This equation indicates that for a constant value of the top quark
Yukawa coupling, larger values of $\tan\beta$ can be achieved in BRpV 
compared with the MSSM (in the MSSM $\cos\chi=1$) when values of
$\cos\chi$
smaller than one are considered (typically $0.87\lsim|c_{\chi}|\lsim
1$). 
The widening of the allowed region $m_{top}-\tan\beta$ in BRpV is also
observed, although less pronounced, if we use the alternative 
definition of $\tan\beta'=v_u/\sqrt{v_d^2+v_3^2}$ where we have
$t'_{\beta}=t_{\beta}c_{\chi}$. The reason is that $b-\tau$ unification
in
BRpV can be achieved at larger values of $t'_{\beta}$, thus lowering 
$h_t$ \cite{YukUnif}.

Now a word about the neutrino mass. The question is whether the values
of $\cos\chi$ we find are compatible with small neutrino masses. The
tau-neutrino neutrino mass is generated in BRpV via mixing with
neutralinos and a weak-scale-type see--saw type mechanism and can be
expressed as
\begin{equation}
m_{\nu_\tau }\approx{{m_Z^2s_{\zeta}^2}\over
{M_{1/2}(1+t_{\beta}^2c_{\chi}^2)}}
\label{NeuTauMasApp}
\end{equation}
where $s_{\zeta}\equiv\sin\zeta$ is another basis independent invariant
which in our basis is equal to 
\begin{equation}
\sin\zeta={{(\mu v_3+\epsilon_3 v_d)}\over{\sqrt{\mu^2+\epsilon_3^2}
\sqrt{v_d^2+v_3^2}}}\,.
\label{zetadef}
\end{equation}
This parameter, which is proportional to the tau--sneutrino VEV in the
basis where the $\epsilon_3$ term is absent from the superpotential,
has to be small in order to have a small neutrino mass. In models with
universality of soft mass parameters at the GUT scale, this parameter is 
naturally small and calculable, since it is generated by radiative
corrections through the RGE's of the soft parameters. It can be shown
that
\begin{equation}
\sin\zeta\approx s_{\chi}c_{\chi}{{\mu't_{\beta}c_{\chi}\Delta B \pm
\Delta m^2}\over{m_{\tilde\nu}^2}}
\label{zetaApp}
\end{equation}
where $\mu'^2=\mu^2+\epsilon_3^2$, $\Delta m^2=m_{H_d}^2-M_{L_3}^2$,
and $\Delta B=B_3-B$ with $B$ and $B_3$ the bilinear soft mass
parameters associated to $\mu$ and $\epsilon_3$, all at the weak
scale. The fraction at the right hand side of eq.~(\ref{zetaApp}),
which we denote as $\delta$, is a good measure of the cancellation
needed in order to have a small neutrino mass. It is approximately
given by
\begin{equation}
\delta\approx{{\sqrt{m_{\nu_{\tau}}M_{1/2}(1+t_{\beta}^2c_{\chi}^2)}}
\over{s_{\chi}c_{\chi}m_Z}}\,.
\label{finetunnig}
\end{equation}
Considering $M_{1/2}=300$ GeV, $\tan\beta=15$, and $\sin\chi=0.3$, the
amount of cancellation necessary to obtain a neutrino mass
$m_{\nu_{\tau}}=0.1$ eV is given by $\delta\approx 10^{-4}$
($\sin\zeta\approx 3\times 10^{-5}$).  We do not think that this is a
fine tuning. For example note that the same amount of cancellation
between VEV's in the MSSM is necessary in $SO(10)$ models where
$\tan\beta$ needs to be higher than 50.

\section{Conclusions}

In summary, in the context of supersymmetric models with universality
of gauge and Yukawa couplings we have studied the predictions for
$m_{top}$, $V_{cb}$ and $\tan\beta$, implied by the
Georgi--Jarlskog--Nanopoulos ans\"atze for fermion mass
matrices. First, we have investigated the impact of the most recent
experimental measurements of the top quark mass and the CKM matrix
element $V_{cb}$ in the MSSM analysis, which we have updated. As it is
well-known, imposing bottom--tau unification at the GUT scale two
solutions are found in the MSSM, characterized by large and low
$\tan\beta$. Requiring in addition the texture constraint for $V_{cb}$
at the GUT scale within $5\%$, the large $\tan\beta$ solution becomes
disallowed even if we accept the experimental measurements for
$m_{top}$ and $V_{cb}$ at $99\%$ c.l. If we relax the level of
validity of the latter condition, the large $\tan\beta$ solution
starts to reappear and it is fully valid when the conditions at the
GUT scale are imposed to within $40\%$. But no intermediate
$\tan\beta$ solutions emerge.

We have also studied the same predictions in the MSSM--BRpV model,
where a bilinear R--Parity violating term is added to the
superpotential. This model is the simplest and most systematic way to
include the effects of R-parity violation. Since no new interactions
are added, its RGE are unchanged with respect to those of the MSSM.
Nevertheless, boundary conditions for Yukawa couplings at the
supersymmetric threshold are different. The allowed region in the
$\tan\beta-m_{top}$ plane in the MSSM--BRpV is slightly larger than in
the MSSM when we impose the GUT conditions for $V_{cb}$ and
bottom--tau unification within $5\%$. This allowed region for the
MSSM--BRpV grows as we relax the texture conditions on $V_{cb}$ at the
GUT scale. When these conditions are imposed within $40\%$, not only
is the large $\tan\beta$ branch recovered as in the MSSM, but also the
full $\tan\beta-m_{top}$ plane including every $\tan\beta$ value
appears. These effects are compatible even with tau neutrino masses as
small as 0.1 eV. Last but not least, such small $\nu_\tau$ values are
not really required by present phenomenology to the extent that the
atmospheric neutrino data allow for alternative explanations involving
sterile neutrinos~\cite{Gonzalez-Garcia:1998dp}, flavour changing
interactions ~\cite{Gonzalez-Garcia:1998hj} or neutrino
decay~\cite{Barger:1999bg}.

\section*{Acknowledgements}

This work was supported by DGICYT grant PB95-1077 and by the EEC 
under the TMR contract ERBFMRX-CT96-0090.
M.A.D. was partly supported by a postdoctoral grant from Ministerio de 
Educaci\'on y Ciencias of Spain, and partly by the
U.S. Department of Energy under contract number DE-FG02-97ER41022.
J.F. was supported by a spanish MEC FPI fellowship.

%\newpage

\end{document}